\documentclass[12pt]{article}
\usepackage{graphicx}
\usepackage[spanish,english]{babel}

\begin{document}

Preprint \hfill   \hbox{\bf SB/F/02-295}
\hrule \vskip 2cm
\pagestyle{empty}
\centerline{ {\large \bf Radiative Processes of the DeWitt-Takagi Detector}}
\vskip 1cm \centerline{D.E.D\'{\i}az and J.Stephany}
\vskip .5cm
\hskip 2.5cm {\small \it Departamento de F\'{\i}sica, Universidad Sim\'on Bol\'{\i}var,}

\hskip 2.7cm {\small \it Apartado Postal 89000, Caracas 1080-A,Venezuela}
\vskip .2cm
\vskip .2cm \hskip 3.7cm {\small \it
{ddiaz@fis.usb.ve,stephany@usb.ve}}

\vskip 2cm
\centerline{\bf Abstract}
\vskip 0.5cm
We examine the excitation of a uniformly accelerated DeWitt-Takagi
detector coupled quadratically to a Majorana-Dirac field. We
obtain the transition probability from the ground state of the
detector and the vacuum state of the field to an excited state
with the emission of a Minkowski pair of quanta, in terms of
elementary processes of absorption and scattering of Rindler
quanta from the Fulling-Davies-Unruh thermal bath in the
co-accelerated frame.

\vskip 1.5cm
\centerline{\bf Abstract}
\vskip 0.5cm
Examinamos la excitaci\'on de un detector de DeWitt-Takagi
uniformemente acelerado, acoplado cuadr\'aticamente al campo de
Majorana-Dirac. Obtenemos las probabilidades de transici\'on desde
el estado base del detector y el vac\'{\i}o del campo al estado
excitado con la emisi\'on de un par de cuantos de Minkowski, en
t\'erminos de procesos elementales de absorci\'on y dispersi\'on
de cuantos de Rindler del ba\~no t\'ermico de Fulling-Davies-Unruh
en el sistema co-acelerado.

\vskip 2.5cm
\hrule
\bigskip
\centerline{\bf UNIVERSIDAD SIMON BOLIVAR} \vskip .5cm \vfill

\section{Introduction}

\hspace{4mm} The introduction of the concept of particle detectors
in quantum field theory in accelerated reference frames and curved
spacetimes provides an operational definition of the concept of
particle and allows the study of particle production in classical
gravitational backgrounds \cite{DeW79},\cite{Bir82}.

In this context, the detectors are idealized point objects with
internal energy levels which follow a classical trajectory  and
are coupled to a quantum field. By computing the transition rates
of the detector one obtains information of the quantum field from
the point of view of non static observers
\cite{DeW79},\cite{Bir82}, \cite{Gin87}. For a uniformly
accelerated detector, this can be done either by quantizing the
field in terms of Minkowski modes or working with the alternative
quantization procedure \cite{Ful73} using Rindler modes . In the
second approach the detector reacts as if immersed in a thermal
bath of Rindler particles, which is called Fulling-Davis-Unruh
bath.

In a previous work \cite{Dia01} we show that for a scalar field
the results of the calculations based on emission/absorption of
Rindler quanta in the presence of the Fulling-Davis-Unruh bath
give in a simple and physically appealing way the same result of
the standard computation involving Minkowski quanta. This was done
by analyzing the response function of the detector and is in
agreement with previous results using different techniques
\cite{Wal84}, \cite{Higu92},\cite{Ren94}. In the limit of a zero
energy gap detector we reobtained the classical result for the
radiation of a charged particle.

In this note we obtain an analogous result for the fermionic case.
In particular we show that the response of a DeWitt-Takagi
detector prepared in the ground state and uniformly accelerated
through the Minkowski vacuum of the massless Dirac field is
exactly what one would expect for a static detector immersed in
the Fulling-Davies-Unruh thermal bath. In this form we get a
direct particle interpretation  in terms of elementary processes
of inelastic scattering and absorption of Rindler quanta for the
emission of a Minkowski pair.

\section{Dirac Field in the Rindler Wedge}

\hspace{4mm} Let us begin by considering the formulation of a massless Dirac
field in Rindler coordinates\cite{Rin77}. The Rindler wedge $R_+$
is the region $x>\left|t\right|$ of Minkowski spacetime, covered
by coordinates $(\zeta,\xi,y,z)$ which are related to Minkowski
coordinates by
\begin{equation}
x=\xi\cosh{\zeta}\,\qquad\qquad\,t=\xi\sinh{\zeta}.
\end{equation}

\begin{figure}
\begin{center}
\resizebox{80mm}{!}{\includegraphics{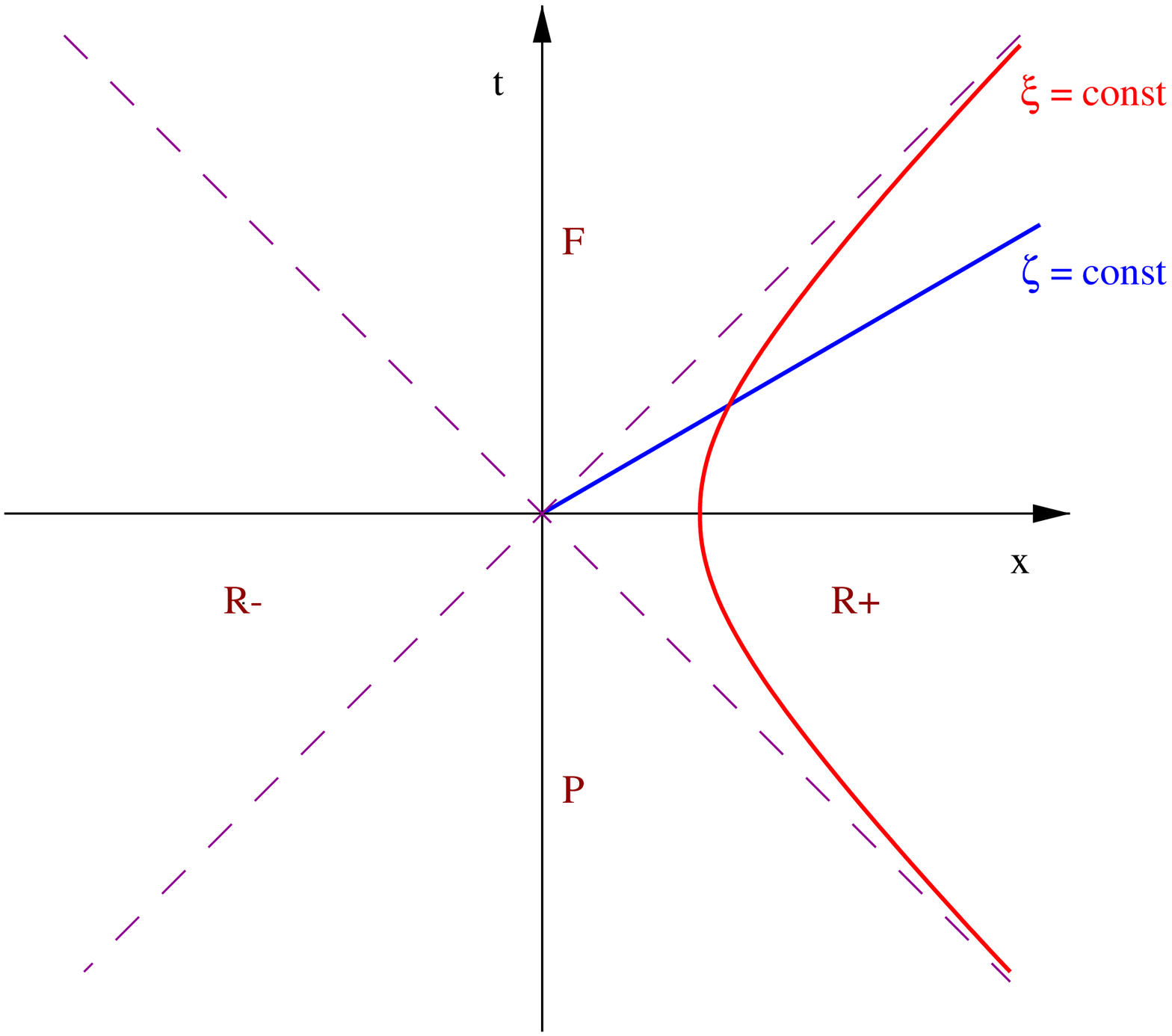}}
\caption{Figure 1. Rindler´s partition of Minkowski space}
\label{fig1}
\end{center}
\end{figure}

The Minkowski line element in these coordinates is
\begin{equation}
 ds^2 = -\xi^2d\zeta^2+d\xi^2+dy^2+dz^2.\end{equation}

Static observers in Rindler coordinates $(\xi$=$const)$ perform
hyperbolic motions in Minkowski spacetime with acceleration
$a={\xi}^{-1}$ and proper time $\tau=\zeta/a$.

In order to write down the solution of the massless Dirac equation
\begin{equation}
 \gamma^{\mu}\nabla_{\mu}\psi=0\end{equation}
in the Rindler wedge, we  follow Candelas and Deutsch approach
\cite{Can78} in the Majorana representation and choose a diagonal
vierbein, which gives only one non vanishing component of the
spinor connection
\begin{equation}
 L^{\mu}_{\alpha}=\mbox{diag}\;[\frac{1}{\xi},1,1,1]\qquad\qquad
 \Gamma_{\zeta}=\frac{1}{2}\gamma^0\gamma^1\,.\end{equation}

The positive frequency solutions of the Dirac equation with
respect to Rindler time are then given by
\begin{equation}\psi(x|\nu,\mathbf{k},\lambda)=\sqrt{\frac{k\cosh{
 \pi\nu}}{2\pi^4}}e^{-i\nu\zeta+\mathbf{kx}}
 K_{i\nu+\frac{1}{2}\gamma^0\gamma^1}(k\xi)\chi(\mathbf{k},\lambda),\end{equation}
in terms of the multicomponent McDonald functions introduced in
Ref.\cite{Can78} and a constant spinor $\chi$. Thus, the field
operator may be decomposed as
\begin{equation}\psi(x)=\sum_{\lambda=\pm} \int_{0}^{\infty} d\nu \int d^{2}
   \mathbf{k}\left\{a(\nu,\mathbf{k},\lambda)\psi(x|\nu,\mathbf{k},
   \lambda)+H.c.\right\}\end{equation}
with the anti-commutation relations on the creation/annihilation
operators.

\section{DeWitt-Takagi Detector}

\hspace{4mm}The conceptually simplest detector one can use to probe the
quantum Dirac field, was proposed by Takagi \cite{Tak86}, and is
obtained by coupling the DeWitt monopole detector to the free
Dirac field by means of the interaction  lagrangian
\begin{equation}
 \mathbf{L}_{int}=m(\tau)\overline{\psi}[x(\tau)]\psi[x(\tau)].
\end{equation}

Assuming that the field is initially in the Minkowski vacuum state
and the detector in the ground state, the transition probability
of the detector to the excited state and to all possible final
states of the quantum field is, to lowest order \cite{Bir82},
\begin{equation}
 \mathcal{P}(E)= |\langle E|m(0)|0\rangle |^2\;\mathcal{F}(E),
\end{equation}
where the response function is given by
\begin{equation}
 \mathcal{F}(E)=\sum_{|f\rangle}\left|\int^\infty_{-\infty}d\tau
 e^{iE\tau}\langle f|
 \overline{\psi}[x(\tau)]\psi[x(\tau)]|0_M\rangle\right|^2.
\end{equation}

The response function is proportional to the proper time interval,
and one can work with the response function per unit proper time
interval $\frac{\mathcal{F}(E)}{T}$ as the relevant physical
quantity. To investigate the contributing terms in the amplitude
above one first expands $\psi$ in Minkowski modes and let it act
on the Minkowski vacuum so that only the creation part of $\psi$
contributes. After that, the action of $\overline{\psi}$ can yield
two possible contributions, the reabsorption of the created
quantum which can be eliminated by normal ordering of the
interaction term, or the creation of another quantum, ending up
with a pair of Minkowski quanta in the final state.

Therefore, to lowest order the only possible final state in terms
of Minkowski quanta is the one with a pair and the response
function per unit proper time interval is proportional the tree
level emission rate of a  Minkowski pair.

\section{Rindler Mode Calculation}

\hspace{4mm}The alternative way to handle this computation is to use the
Rindler mode expansion of the field. After performing one time
integral one gets for the response function
$\frac{\mathcal{F}(E)}{T}$
\[
\sum_{|f\rangle}\sum_{J,J'}\delta(E+a\nu'-a\nu)\;\left|\langle
f|\,a^{\dag}_{J'}\,a_{J}\,|0_M\rangle
   \,\overline{\psi}_{J'}(\mathbf{x_0})\,\psi_J(\mathbf{x_0})\right|^2\]
\vspace{-5mm}
\[+\,\delta(E-a\nu'-a\nu)\;\left|\langle
f|\,a_{J'}\,a_{J}\,|0_M\rangle
 \,\overline{\psi}^*_{J'}(\mathbf{x_0})\,\psi_J(\mathbf{x_0})\right|^2
 \]
\vspace{-5mm}
\begin{equation}+\,\delta(E-a\nu'+a\nu)\;\left|\langle
f|\,a_{J'}\,a^{\dag}_{J}\,|
   0_M\rangle\,\overline{\psi}^*_{J'}(\mathbf{x_0})\,\psi^*_J(\mathbf{x_0})
   \right|^2,\end{equation}
where we are using the compact notation
$J\equiv\{\nu,\mathbf{k},\lambda\}$ and $\psi_J(\mathbf{x_0})$
stands for the spatial part of $\psi(x|\nu,\mathbf{k},\lambda)$ at
$\xi=a^{-1},y=z=0$.

Using the completeness of the final states, the fact that the
restriction of the Minkowski vacuum state
to the Rindler wedge gives the Fermi-Dirac factor for the
expectation value of Rindler quanta number and the
anti-commutative nature of the fermionic operators, we can
identify the contributing terms in the above expression. These are
the following: annihilation of a Rindler quantum with frequency
$a\nu$ with the creation of another with frequency $a\nu'=a\nu-E$,
annihilation of a pair with frequencies $a\nu+a\nu'=E$ and
finally, creation of an anti-particle with frequency $a\nu$ and
the annihilation of another with frequency $a\nu'=a\nu+E$. In this
case the particle and anti-particle are the same, but the same
result holds also for the full electron-positron field.

The tree level emission rate is then obtained in the
co-accelerating frame as the contribution of three elementary
processes, each one weighted with the appropriate statistical
factor and satisfying the energy balance conditions mentioned above
\[
   \frac{d\mathcal{P}_{emis,e^-_{M}e^+_{M}}}{d\tau}= \sum_{J,J'}
   \left[\frac{1}{e^{2\pi\nu}+1}\right]\left[\frac{1}{e^{2\pi\nu'}+1}
  \right]\frac{d\mathcal{P}_{absor,e^-_{R}e^+_{R}
  }}{d\tau}\Big|_{J,J'}
\]
\vspace{-5mm}
\[+
\left[\frac{1}{e^{2\pi\nu}+1}\right]\left[1-\frac{1}{e^{2\pi\nu'}+1}
  \right]\frac{d\mathcal{P}_{disp,e^-_{R}}
  }{d\tau}\Big|_{J,J'}\]
\vspace{-5mm}
\begin{equation}
+\left[1-\frac{1}{e^{2\pi\nu}+1}\right]\left[\frac{1}{e^{2\pi\nu'}+1}
  \right]\frac{d\mathcal{P}_{disp,e^+_{R}}
  }{d\tau}\Big|_{J,J'}
\end{equation}

There is induced absorption and, in contrast to the bosonic case,
the emission is attenuated rather than stimulated. This is a
manifestation of Pauli's exclusion principle. It is worth
mentioning that a similar attenuation effect was obtained by
L. Parker\cite{Par71} in the context of fermion production in an
expanding gravitational background.

\begin{figure}
\begin{center}
\resizebox{80mm}{!}{\includegraphics{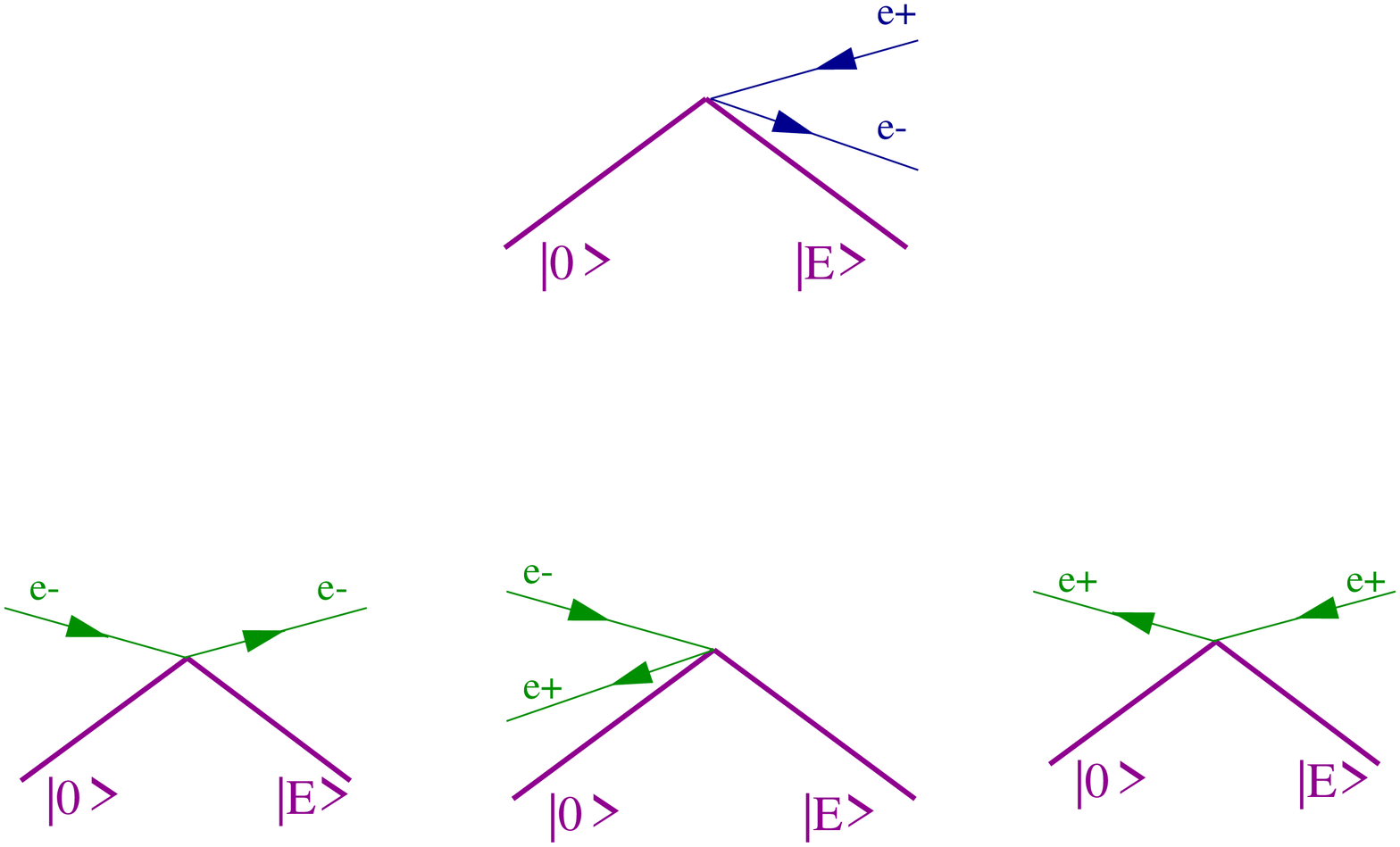}}
\caption{Figure 2. Minkowski pair emission, equivalent to absorption/scattering
 of Rindler quanta}
\label{fig2}
\end{center}
\end{figure}

 A graphical representation of the two different
particle interpretations of the 'click' of the DeWitt-Takagi
detector is given in fig.2.

We note that Matsas and Vanzella \cite{Mat99}, studying the
influence of acceleration in particle decays, have proposed  to
describe the proton and neutron states of the nucleon with a
semi-classical current coupled to the electron and neutrino
fields. This is  a slight modification of the DeWitt-Takagi
detector and therefore in the accelerated frame there are also
three contributing processes to the proton decay rate and the
branching ratios for each channel can be calculated.

\section{Conclusion}

\hspace{4mm} We discussed the particle interpretation of the
behavior of a monopole detector in hyperbolic motion  coupled to a
fermionic field. For an inertial observer the excitation of the
detector comes with the emission of a Minkowski pair of particles
at a rate which from the point of view of an accelerated observer
corresponds to the combined effects of absorption and scattering
of Rindler quanta from the Fulling-Davies-Unruh thermal bath with
the appropriate statistical weights. The presence of fermions in
the final state results in an attenuation of the emission rate of
Rindler quanta, in agreement with similar effects reported in the
literature. The equivalence found here may prove relevant in the
analysis of fermion absorption and emission on more general
backgrounds, e.g. the exterior region of a black hole.


\end{document}